\documentclass[pre,aps,twocolumn,showpacs]{revtex4}
\usepackage{epsfig}
\include{graphics}
\begin{document}
\input{epsf}
\title{A generalized Cahn-Hilliard equation for biological applications}

\author{Evgeniy Khain$^{1}$ and Leonard M. Sander$^{2}$}

\affiliation{$^{1}$ Department of Physics, Oakland University,
Rochester, Michigan 48309}

\affiliation{$^{2}$ Department of Physics and Michigan Center for
Theoretical Physics, The University of Michigan, Ann Arbor,
Michigan 48109}

\begin{abstract}
Recently we considered a stochastic discrete model which describes
fronts of cells invading a wound \cite{KSS}. In the model cells
can move, proliferate, and experience cell-cell adhesion. In this
work we focus on a continuum description of this phenomenon by
means of a generalized  Cahn-Hilliard equation (GCH) with a
proliferation term. As in the discrete model, there are two
interesting regimes. For subcritical adhesion, there are
propagating "pulled" fronts, similarly to those of
Fisher-Kolmogorov
equation. The problem of front velocity
selection is examined, and our theoretical predictions are in a
good agreement with a numerical solution of
the GCH
equation. For supercritical adhesion, there is a nontrivial
transient behavior, where density profile exhibits a secondary
peak. To analyze this regime, we investigated
relaxation
dynamics for the Cahn-Hilliard equation without proliferation. We found
that the relaxation process exhibits self-similar behavior. The
results of continuum and discrete models are in a good agreement
with each other for the different regimes we analyzed.
\end{abstract}
\maketitle

\section{Introduction}

In this paper we propose a continuum method for dealing with cells
that move, proliferate and interact via adhesion. This problem
arises in models for wound healing \cite{wound} and tumor growth
\cite{tumor}. It is easy to formulate a discrete model for these
processes \cite{KSS}. However, proceeding to the continuum limit is
non-trivial \cite{Turner}.

Consider first a simple discrete model for diffusion and
proliferation. Each lattice site can be empty or once occupied. At
each time step, a particle is picked at random. Then it can either
jump to a neighboring empty site, or proliferate there (a new
particle is born). We can ask for is the continuum analog of this
model? It was shown \cite{FKPP1,FKPP,FKPP2} that for {\it small}
proliferation rates the propagating fronts in this discrete system
can be described by the Fisher-Kolmogorov equation \cite{fisher1}
(FK):
\begin{equation}
\frac{\partial u}{\partial t} = \bar{D}\,\frac{\partial^2
u}{\partial x^2} + \alpha\,u\,(1-u), \label{fisher_eq}
\end{equation}
where $u$ is the (local) cell density, $\bar{D}$ is the cell
diffusion coefficient, and $\alpha$ is the rate of proliferation
\cite{fisher2}. Eq.~(\ref{fisher_eq}) admits solutions in a form
of propagating fronts, but the velocity selection is a nontrivial
problem. There is a range of possible velocities; initially
sufficiently localized density profiles develop into propagating
fronts with the ``critical" velocity, $v =
2\sqrt{\bar{D}\,\alpha}$ \cite{Aronson}.

Next consider a discrete model which includes diffusion and
cell-cell adhesion, but not proliferation. When a particle is
picked at random, the probability to jump decreases with the
number of nearest neighbors, to take into account cell-cell
adhesion \cite{KSS}. This scheme can be mapped into the
Ising model \cite{Onsager,Ising}, by identifying an empty site
with spin down, and an occupied site with spin up. There is a
simple relation between the average density, $u$, and the average
magnetization, $m$, in the Ising model: $u=(m+1)/2$. The number of
particles is fixed because there is no proliferation, and there
are nearest-neighbor interactions between particles. Above the
critical strength of cell-cell adhesion (or below a critical
temperature in the Ising model), the homogeneous state becomes
unstable, which leads to phase separation between high density
clusters and a dilute gas of particles. The dynamics of phase
separation and coarsening (where larger clusters grow at the
expense of smaller ones) is usually described by the Cahn-Hilliard
equation \cite{Cahn}, a version of this equation can be derived
directly from the microscopic model \cite{review}.

We can easily add proliferation to this lattice model
\cite{KSS}, so that we have diffusion, proliferation, and
cell-cell adhesion. In this paper we suggest that the proper
candidate for a continuum description is a Cahn-Hilliard equation
with a proliferation term added, the GCH equation.

The rest of the paper is organized as follows. In Sec. 2 we
present both discrete and continuum models which include
diffusion, proliferation, and cell-cell adhesion and present a
general phase diagram. Section 3 describes the front propagation
problem for subcritical adhesion. Section 4 focuses on a
supercritical adhesion both for zero and nonzero proliferation.
Section 5 includes a brief discussion and summary of our results.

\section{Discrete and continuum models}

\subsection{Discrete model}

We review the discrete model for diffusion, proliferation, and
cell-cell adhesion \cite{KSS}. Consider a square
two-dimensional lattice in a channel geometry. The lattice
distance is assumed to be equal to cell diameter, taking into
account hard-core exclusion. Initially, we put cells into the left
part of the channel. We take $x$ to measure distance along the
channel. A cell is picked at random, and one of the four
neighboring sites is also picked at random. If this site is empty,
the cell can proliferate to this site (so that a new cell is born
there), or migrate there. We denote the probability for
proliferation by $\alpha$. Cell-cell adhesion is represented by a
probability for migration that decreases with the number of
nearest neighbors: $p_{migr} = (1-\alpha)(1-q)^n$, where $0 \leq q
< 1$ is the adhesion parameter, and $1 \leq n \leq 4$ is the
number of nearest neighbors. The case $q=0$ means no adhesion and
reduces to the model of Refs. \cite{FKPP1,FKPP,FKPP2}. For nonzero
$q$, it is much harder to a cell to diffuse if it has many
neighbors. After each step  time is advanced by $1/N$, where $N$
is the current number of cells.

\emph{Without} proliferation the model can be mapped into the
Ising model, as we pointed out in \cite{KSS}. In this
mapping the adhesion parameter $q$ is identified with
$1-\exp(-J/k_BT)$ where $T$ is the temperature, $k_B$ is
Boltzmann's constant, and $J$ is the coupling strength in the
magnetic model, and the average density $u$ is identified with
$(m+1)/2$, where $m$ is the average magnetization. The mapping is
possible because our dynamical rules satisfy detailed balance.
Therefore, the statics of our model is the same as in the Ising
model. By statics, we mean a phase diagram $(m, T)$ (or $(u, q)$
in our case) which has stable and unstable regions. In the stable
region, a homogeneous state (with uniformly distributed cells)
remains homogeneous; in contrast, in the unstable region phase
separation occurs and large clusters are formed.

\begin{figure}[ht]
\centerline{\includegraphics[width=3.0in,clip=]{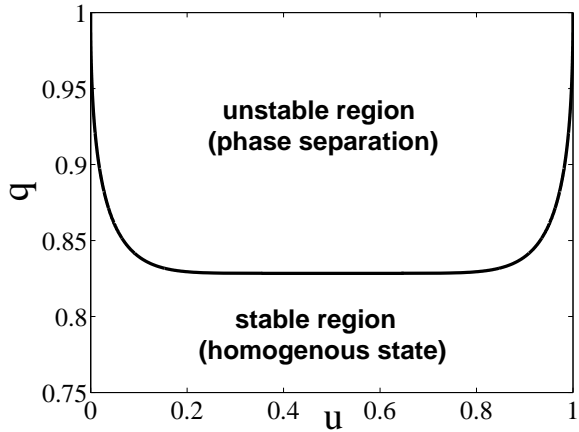}}
\caption{Phase diagram without proliferation. The critical
adhesion parameter as a function of density as given by
Eq.~(\ref{Onsager}) is shown by solid line. This curve separates
two qualitatively different regions. In the stable region, a
homogeneous state (with uniformly distributed cells) remains
homogeneous; in contrast, in the unstable region phase separation
occurs and large clusters are formed.} \label{threshold}
\end{figure}

The two-dimensional Ising model was solved by Onsager
\cite{Onsager}, and the curve $m(T)$, which separates the stable
and unstable regions, is known (see Fig.~\ref{threshold}). In our
case the threshold is given by the critical adhesion parameter
$q_c$ as a function of average density $u$:
\begin{equation}
u = \frac{1}{2} \pm \frac{1}{2}\left[ 1 -
\frac{16(1-q_{c})^2}{q_{c}^4} \right]^{1/8}. \label{Onsager}
\end{equation}
The unstable region corresponds to $q>q_c$, so for supercritical
adhesion there is a phase separation and large clusters are
formed. Interestingly, even if we start with $q<q_c$, the
initially homogeneous state can become unstable when one turns on
proliferation, which leads to phase separation and clustering.

\subsection{Continuum approach}

To describe the coarse-grained dynamics of the discrete model, we
take a continuum approach. A model equation, which describes the
dynamics of phase separation with conserved order parameter
(without proliferation) is the Cahn-Hilliard equation
\cite{Cahn,review}. We now formulate the GCH equation by
adding a proliferation term:
\begin{equation}
\frac{\partial u}{\partial t} = \frac{\partial^2}{\partial x^2}
\left(\ln(1-q)\frac{\partial^2 u}{\partial x^2}
+\frac{df}{du}\right) + \alpha\,u\,(1-u) \label{Cahn}
\end{equation}
where $u$ is the local density, $f$ is the local free energy, and
$\alpha$ is the rate of proliferation. The gradient term in the
total free energy functional is given by $(1/2)\,J\,(\partial
c/\partial x)^2$, where $J$ represents interatomic interactions
(for example, the coupling strength in the Ising model). This
leads (in dimensionless form) to the $\partial^2/\partial x^2
[-(J/kT)\,(\partial^2 u/\partial x^2)]$ term in Eq.~(\ref{Cahn}).
The mapping $q=1-\exp(-J/k_BT)$ explains the $\ln(1-q)$
coefficient in Eq.~(\ref{Cahn}). Usually, the mean field form of
the local free energy is assumed:
\begin{equation}
f(u) = 0.5\,a\,(u-0.5)^2 + 0.25\,b\,(u-0.5)^4. \label{freeenergy}
\end{equation}

\begin{figure}[ht]
\centerline{\includegraphics[width=3.0in,clip=]{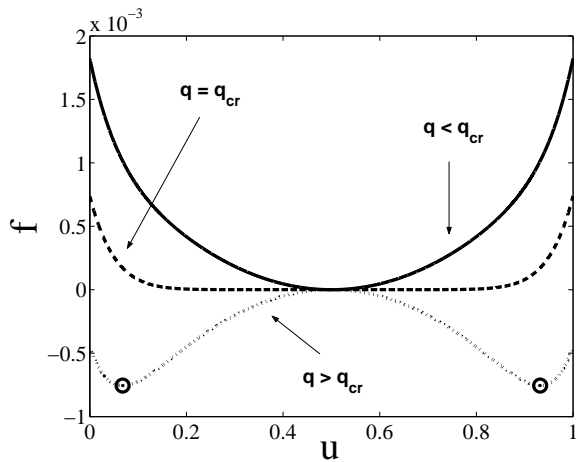}}
\caption{The local free energy for subcritical (solid line,
$q=0.81$), critical (dashed line), and supercritical (dotted line,
$q=0.85$) adhesion. The minimal adhesion threshold is given by
$16(1-q_{cr})^2/q_{cr}^4=1$, which gives $q_{cr}=0.8284...$ Two
circles denote two stable phases, see Eq.~(\ref{Onsager}).}
\label{free_energy}
\end{figure}

Figure \ref{free_energy} shows the local free energy both for
subcritical and supercritical adhesion. The only extremum
(minimum) of $f(u)$ for subcritical adhesion is at $u=1/2$, so the
homogeneous state is stable. For supercritical adhesion, the
extremum at $u=1/2$ becomes a maximum, the homogeneous state is no
longer  stable, and two new minima appear, with the densities given
by Eq.~(\ref{Onsager}).

The constants $a$ and $b$ in Eq.~(\ref{freeenergy}) are chosen in
such a way that the free energy functional satisfies several
important conditions. First, we demand that the phase transition
threshold is exact (and not its mean field approximation) as given
by Eq.~(\ref{Onsager}). Second, as the adhesion parameter $q$
tends to zero, $b$ should go to zero. Then Eq.~(\ref{Cahn})
transforms into the FK equation. In addition, the diffusion
coefficient in Eq.~(\ref{fisher_eq}) should be $\bar{D}=1/4$ (as
can be derived from the discrete model without adhesion
\cite{FKPP2}), so that $\lim_{q \longrightarrow 0} a = 1/4$. We
chose the following expressions:
\begin{eqnarray}
a &=&
-\frac{q-q_{cr}}{|q-q_{cr}|^{3/4}}\,\frac{c(q)}{4q_{cr}^{1/4}},
\nonumber
\\
b &=& \left[\frac{q-q_{cr}}{1 -
16(1-q)^2/q^4}\right]^{1/4}\,\frac{c(q)}{q_{cr}^{1/4}},
\label{constants}
\end{eqnarray}
where the only restriction on the function $c(q)$ is that it
should tend to unity when $q$ goes to zero. This function will be
used to fit the continuum results with results of discrete
simulations. Note that the theoretical analysis is performed for
the general form of local free energy and the specific relations
(\ref{constants}) are used only when comparing theoretical
predictions with numerical simulations.

In the next section we consider the regime of subcritical adhesion
and focus on front propagation.

\section{Subcritical adhesion: front propagation}

\subsection{Theory}

Initially, we put cells into the left part of the channel; $x$
measures the distance along the channel. In the initial state all
sites with $x<0$ are occupied and the rest empty. For $t>0$ cells
diffuse and proliferate along the channel and form an advancing
front. To analyze those fronts, we look for the solutions in the
form $u=u(\xi = x-vt)$ in Eq.~(\ref{Cahn}), where the front
velocity $v$ is {\it unknown}. This gives
\begin{equation}
\ln(1-q)u^{\prime\prime\prime\prime} +
\frac{d^2f}{du^2}u^{\prime\prime} +
\frac{d^3f}{du^3}{u^{\prime}}^2 + vu^{\prime} + \alpha u (1-u) =
0. \label{Cahn2}
\end{equation}

In order to understand velocity selection, we linearize
Eq.~(\ref{Cahn2}) in the tail region $u=0$, in a similar way to
the analysis of pulled fronts in the FK equation. Substituting $u
\propto \exp(\lambda \xi)$, we find
\begin{equation}
E = \ln(1-q)\lambda^4 +
\left(\frac{d^2f}{du^2}|_{u=0}\right)\lambda^2 + v\lambda + \alpha
= 0. \label{polinom}
\end{equation}
The behavior of the density front in the tail region depends on
the sign of the determinant
$$D(v) = -\left(\frac{P^2}{9} + \frac{4r}{3}\right)^3 + \left(-\frac{P^3}{27} +
\frac{4Pr}{3} - \frac{Q^2}{2}\right)^2,$$ where
\begin{eqnarray}
P&=&\left(\frac{d^2f}{du^2}|_{u=0}\right)\frac{1}{\ln(1-q)},
\nonumber
\\
r&=& \frac{\alpha}{\ln(1-q)}, \,\,\,\,\,\mbox{and}\,\,\,\,\,Q =
\frac{v}{\ln(1-q)}. \nonumber
\end{eqnarray}
As in the FK equation, there is an interval of possible
velocities, $v>v_{min}$. We checked numerically that for small
enough $\alpha$ (or small enough $q$, see below), velocity
selection is determined by the condition $D=0$. This gives the
minimum velocity of front propagation (the critical velocity):
\begin{eqnarray}
v_{cr}^2 = \frac{8\alpha}{3}\left(\frac{d^2f}{du^2}|_{u=0}\right)
-
\frac{2}{27\ln(1-q)}\left(\frac{d^2f}{du^2}|_{u=0}\right)^3\,\times
\nonumber
\\
\left[1 -
\left(1+12\alpha\ln(1-q)\left(\frac{d^2f}{du^2}|_{u=0}\right)^{-2}\right)^{3/2}\right].
\label{velocity}
\end{eqnarray}
As expected, $v_{cr}$ tends to zero when $\alpha$ goes to zero
(fronts do not propagate for zero proliferation), and $v_{cr}$
tends to the value of the FK equation with $\bar{D}=1/4$ [see
Eq.~(\ref{fisher_eq})], $v_{cr}\longrightarrow\sqrt\alpha$, when
the adhesion parameter $q$ goes to zero.

\begin{figure}[ht]
\centerline{\includegraphics[width=3.0in,clip=]{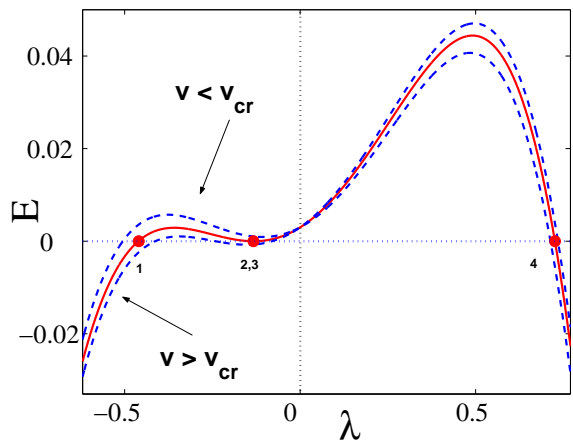}}
\caption{The characteristic equation $E$ [Eq.~(\ref{polinom})] for
different values of front velocity. The solid line corresponds to
the critical velocity [given by Eq.~(\ref{velocity})], the two
dashed lines correspond to larger and smaller velocities. The
intersection of the horizontal dotted line $E=0$ with the solid
line gives four roots of the characteristic equation (circles).
The parameters are $q = 0.4, \alpha = 0.003$.} \label{polinom3}
\end{figure}

The selection rule is illustrated in Fig.~\ref{polinom3}. If the
velocity is slightly larger than $v_{cr}$, Eq.~(\ref{polinom}) has
four real roots: three negative and one positive. As the velocity
approaches $v_{cr}$, two negative roots approach each other and
coincide exactly when $D=0$, see Fig.~\ref{polinom3}. It is reasonable to suppose,
in view of the results for the FK equation, that this is the
selected velocity of front propagation; see \cite{saarloos}.
For smaller velocity these
two roots become complex, which is not allowed as it results in an
oscillatory behavior of the density in the tail region.

\begin{figure}[ht]
\centerline{\includegraphics[width=3.0in,clip=]{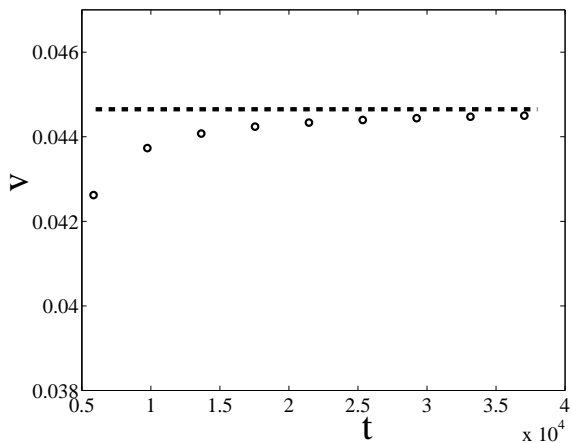}}
\caption{Front velocity from numerical solution of
Eq.~(\ref{Cahn}) as a function of time. The velocity slowly
approaches the theoretical value, similarly to the situation in FK
equation. The parameters are $q = 0.6, \alpha = 0.003$.}
\label{vel}
\end{figure}

To test these predictions, we performed numerical simulations of
the Eq.~(\ref{Cahn}). We used the third order Runge-Kutta method,
with a mesh size $\delta x = 1.0$, and a time step $\delta t =
0.03$. Initial conditions were localized: $u=0$ ahead of the
front, $u=1$ behind the front, and the interface had the form
$u=\exp(-x)/[1+\exp(-x)]$. Simulations confirm that the front
velocity tends to the value given by Eq.~(\ref{velocity}). As in
the FK equation, $v$ approaches $v_{cr}$ quite slowly, see
Fig.~\ref{vel}.

Note that Eq.~(\ref{velocity}) becomes invalid when $\alpha$ (or
$q$) are large enough, so that
$1+12\alpha\ln(1-q)(d^2f/du^2|_{u=0})^{-2}$ becomes negative. In
this case the characteristic equation (\ref{polinom}) has two real
roots [one negative ($\lambda_1$) and one positive ($\lambda_4$)]
and two complex conjugate roots with negative real part
($\lambda_{2,3} = \gamma \pm i\omega$). Therefore, the density in
the tail region is oscillatory and becomes negative, which is
forbidden. This is an inherent feature of the time-dependent
Eq.~(\ref{Cahn}). It is known that solutions of fourth-order
differential equations generally do not remain positive
\cite{Smereka}. A similar effect occurs in the extended
Fisher-Kolmogorov equation (EFK) \cite{saarloos}. It was shown
that in some region of parameters localized initial conditions can
not develop into a uniformly translating front solution. Instead,
for sufficiently sharp initial conditions one has an {\it envelope
front}: a moving front creates a periodic array of kinks behind it
\cite{saarloos}. These oscillations between $u=1$ and $u=-1$ occur
due to the cubic nonlinearity in the reaction term of the EFK
equation. In our case, $u=-1$ is not a fixed point, so negative
perturbation are not stabilized (in fact, numerical simulations of
Eq.~(\ref{Cahn}) show a finite-time explosion in this region of
parameters.) In addition, $u<0$ has no physical meaning in our
case. A possible way to overcome this problem is to demand $u \geq
0$ when solving Eq.~(\ref{Cahn}). In this case, the density
profile is not analytic, as in problems with compact support: the
density becomes zero at some $\xi_{crit}$ and remains zero for
$\xi>\xi_{crit}$. This profile propagates with a well defined
velocity, see Fig.~\ref{comparison}. We believe that this is the
only type of solution which can be chosen in this regime.

\subsection{Comparison with discrete simulations}

\begin{figure}[ht]
\centerline{\includegraphics[width=3.0in,clip=]{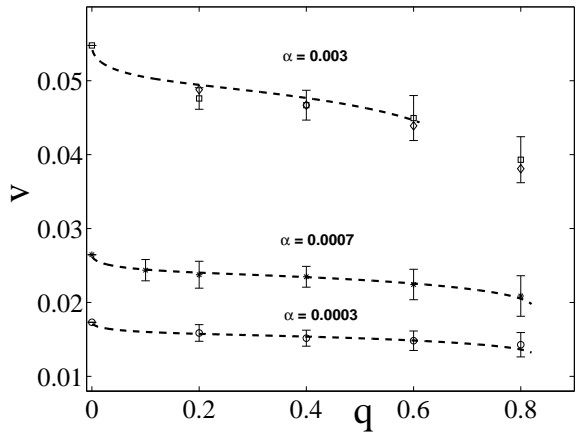}}
\caption{Front velocity as a function of the adhesion parameter
$q$ for different values of proliferation $\alpha$. The
theoretical predictions are given by Eq.~(\ref{velocity}) (dashed
lines). The front velocity in the discrete system is obtained by
averaging over many realization. The calculations in
Eq.~(\ref{velocity}) were performed for $c(q) = 1 - 0.75q^{1/2}$
in the expression for free energy, which gives the best agreement
with discrete simulations. One can see that the theoretical curve
corresponding to large $\alpha$ becomes invalid for large $q$,
which is related to the oscillatory behavior of density tails.
Nevertheless, the numerically calculated front velocities in this
region are well-defined and agree with those from discrete
simulations. Front velocities computed from the time-dependent
Eq.~(\ref{Cahn}) are shown by diamonds. The values of
proliferation are $\alpha = 0.0003$ (circles), $\alpha = 0.0007$
(asterisks), and $\alpha = 0.003$ (squares).} \label{comparison}
\end{figure}

Now we compare the results of deterministic continuum approach
with stochastic discrete modeling. Figure~\ref{comparison} shows
the front velocity as a function of the adhesion parameter $q$ for
different values of proliferation $\alpha$. The theoretical
predictions are given by Eq.~(\ref{velocity}). The front velocity
in the discrete system is obtained by averaging over many
realizations. One can see an excellent agreement over wide range
of parameters. (Front velocity computed numerically from
Eq.~(\ref{Cahn}) also approaches the same values, see
Fig.~\ref{vel}). The theoretical curve corresponding to large
$\alpha$ becomes invalid for large $q$, which is related to the
oscillatory behavior of density tails. Nevertheless, the
numerically calculated front velocities in this region are
well-defined and agree with those from discrete simulations. This
shows that our theoretical understanding in this case is
incomplete.

\begin{figure}[ht]
\centerline{\includegraphics[width=3.0in,clip=]{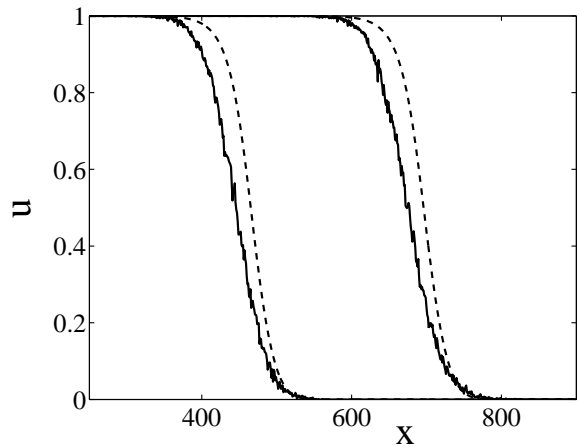}}
\caption{Time series of density profiles in discrete (solid lines)
and continuum (dashed lines) models. The parameters are: $q=0.4$,
$\alpha = 0.003$. The discrete fronts correspond to times $t_{d1}
= 10000$ and $t_{d2} = 15000$, the density profiles computed from
numerical solution of Eq.~(\ref{Cahn}) correspond to times $t_{c1}
= 6000$ and $t_{c2} = 11000$.} \label{profiles}
\end{figure}

Figure~\ref{profiles} shows an example of the corresponding
density profiles from discrete and continuum simulations. The form
of the fronts is very similar, and discrete and continuum fronts
propagate with the same velocity. Note however, that the transient
regime for discrete front is longer.

\section{Supercritical adhesion}

The situation for $q>q_c$ is more complicated. As before, we start
with a sharp front: $u=1$ for $x<0$ and $u=0$ for $x>0$. It turns
out there is a nontrivial and long-lived transient behavior
\cite{KSS}, which we analyze using discrete and continuum
approaches.

\subsection{Nonzero proliferation}

\begin{figure}[ht]
\vspace{0.3cm}
\centerline{\includegraphics[width=3.0in,clip=]{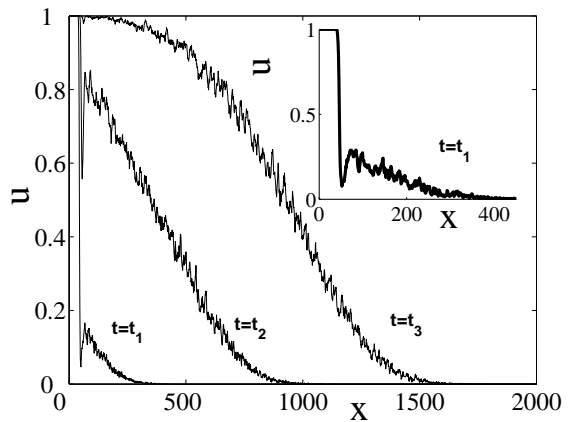}}
\caption{Time series of density profiles in discrete simulations.
An inset shows a magnified picture of the density profile for
early time $t=t_1$. The parameters are: $q = 0.9$, $\alpha =
0.0003$, $t_1 =2.2\times 10^4$, $t_2 =6.6\times 10^4$, $t_3
=11\times 10^4$.} \label{density_sup}
\end{figure}

\begin{figure}[ht]
\vspace{0.3cm}
\centerline{\includegraphics[width=3.0in,clip=]{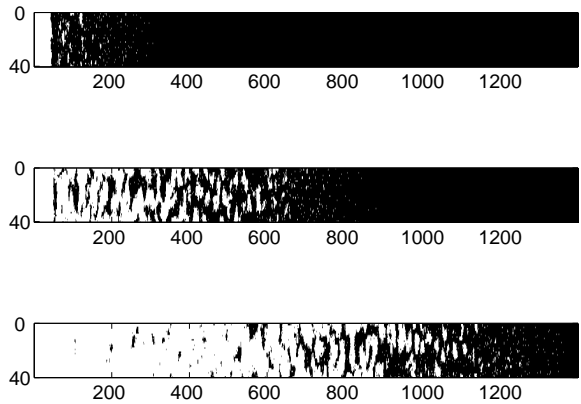}}
\caption{Time series of snapshots, corresponding to the density
profiles in Fig.~\ref{density_sup}. The upper panel corresponds to
$t_1 =2.2\times 10^4$, the middle panel to $t_2 =6.6\times 10^4$,
and the lower panel to $t_3 =11\times 10^4$. The parameters are
the same as in Fig.~\ref{density_sup}. } \label{discrete}
\end{figure}

\begin{figure}[ht]
\vspace{0.3cm}
\centerline{\includegraphics[width=3.0in,clip=]{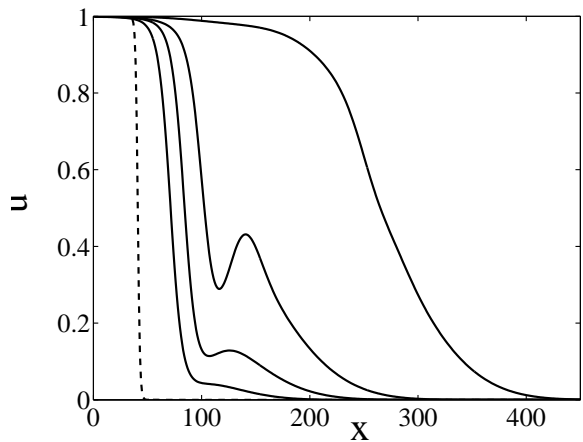}}
\caption{Time series of density profiles from numerical solution
of Eq.~(\ref{Cahn}). The parameters are: $q = 0.9$, $\alpha =
0.0003$, $t_0 = 0$ (dashed line), $t_1 =1.8\times 10^4$, $t_2
=2.4\times 10^4$, $t_3 =3.0\times 10^4$, and $t_4 =4.4\times 10^4$
(solid lines).} \label{cahn_q9a0003}
\end{figure}

We first consider $\alpha \neq 0$, $q>q_c$.
Figure~\ref{density_sup} shows a time series of density profile.
To obtain the profiles, we averaged the density over the channel
width and over many realizations. One realization is shown in
Fig.~\ref{discrete}. A long-lived transient occurs before the
propagating front is formed. An inset shows a magnified picture of
the density profile for early time. An interesting feature is the
secondary density peak. This peak occurs due to phase separation
and cluster formation in the low-density invasive region. At later
times the main front builds up and catches the isolated clusters.
The same feature is present in the continuum approach:
Fig.~\ref{cahn_q9a0003} shows a time series of density profiles
from numerical solutions of Eq.~(\ref{Cahn}).

The transient behavior with a secondary density peak occurs due to
the slow (nonexponential) decay of the tail, which occurs only for
supercritical adhesion. For $\alpha>0$, there are two competing
processes: the (slow) propagation of the front and relatively fast
formation of the secondary peak on the slowly decaying tail. This
slowly decaying tail also exists for zero proliferation. In order
to gain some insights into its formation, we now study the
relaxation dynamics in Cahn-Hilliard equation [Eq.~(\ref{Cahn})]
without proliferation.

\subsection{Zero proliferation}

In this section we focus on the relaxation dynamics (for zero
proliferation) both in the discrete and continuum models.

We start with the discrete lattice model. Initially, all channel
sites with $x<0$ are occupied $(u=1)$, and sites with $x>0$ are
empty $(u=0)$. However, since $q>q_c$, the final state consists of
two phases: a high density phase $u=u_1(q)$ for $x<0$ and a low
density phase $u=u_2(q)$ for $x>0$, see Eq.~(\ref{Onsager}).

\begin{figure}[ht]
\vspace{0.3cm}
\centerline{\includegraphics[width=3.0in,clip=]{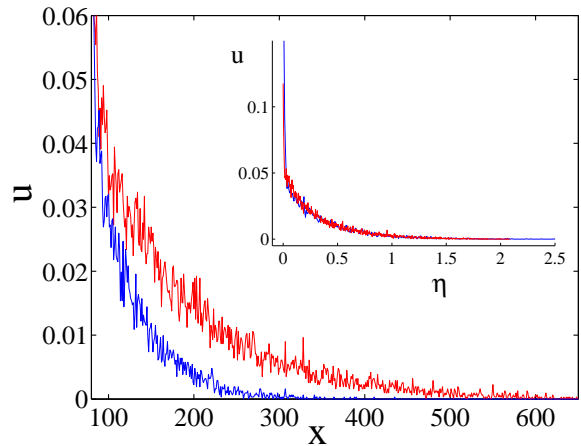}}
\caption{Density profiles (tails) for different times as computed
from discrete model. An inset shows the same density tails as a
function of $\eta = x/\sqrt{t}$. The parameters are $q=0.85$, $t_1
= 2\times 10^4$, $t_2 = 10^5$.} \label{relaxation_d}
\end{figure}

Figure~\ref{relaxation_d} shows the tails of two density profiles
calculated from the discrete model. As before, we averaged over
the channel width and over many realizations. The relaxation
dynamics is self-similar. An inset shows that the same density
tails coincide when measured as a function of $\eta = x/\sqrt{t}$,
as expected for purely diffusive behavior.

The same relaxation dynamics occurs in the continuum model. This
behavior can be easily explained. Considering the small-density
region (the tail), we can neglect the $\ln(1-q)(\partial^2
u/\partial x^2)$ term in Eq.~(\ref{Cahn}). Then, substituting
$u=u(\eta)$ into Eq.~(\ref{Cahn}), we have:
\begin{equation}
\left(\frac{d^2f}{du^2}\right)^{\prime\prime} +
\frac{1}{2}\eta\,u^{\prime}=0, \label{self1}
\end{equation}
where $^\prime$ is the derivative with respect to $\eta$. To solve
this equation we need to specify two boundary conditions at $\eta
= 0$. The first one is just $u(\eta=0) = u_2$ (the low density
stable phase). Then we find $u^{\prime}(\eta=0)$ by a shooting
procedure, demanding $u(\eta\longrightarrow\infty)\longrightarrow
0$.

\begin{figure}[ht]
\vspace{0.3cm}
\centerline{\includegraphics[width=3.0in,clip=]{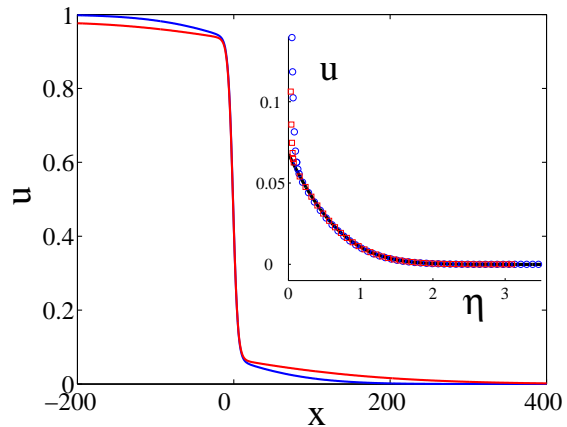}}
\caption{Density profiles for different times (blue curve at $t_1
= 1.5\times 10^4$, red curve at $t_2 = 6\times10^4$) as computed
from Eq.~(\ref{Cahn}). An inset shows that the tails of density
profiles corresponding to different times coincide when measured
as a function of $\eta = x/\sqrt{t}$. Blue circles correspond to
$t_1$, red squares correspond to $t_2$. The asymptotics computed
from Eq.~(\ref{self1}) is shown by the solid black line. The
adhesion parameter is $q=0.85$. The simulations were performed for
$c=4\,q_c^{1/4}$ in the expression for free energy.}
\label{relaxation_c}
\end{figure}

Figure~\ref{relaxation_c} shows this self-similar relaxation
dynamics. Two density profiles depicted in
Fig.~\ref{relaxation_c}, are calculated from the time-dependent
equation (\ref{Cahn}). An inset shows that these profiles (the
tails) corresponding to different times coincide when measured as
a function of $\eta = x/\sqrt{t}$. The asymptotics computed from
Eq.~(\ref{self1}) is in an excellent agreement with numerical
simulations.

This self-similar behavior means that the decay rate is inversely
proportional to $t^{1/2}$, so it becomes lower with time. This is
the base of the formation of secondary density peak in the
transient regime for nonzero proliferation.

\section{Summary and discussion}

In this work we formulated and examined a continuum model for
motile and proliferative cells, which experience cell-cell
adhesion. We describe collective behavior of the cells by using a
modified Cahn-Hilliard equation with an additional proliferation
term. We identified and analyzed different parameter regimes:
front propagation for subcritical adhesion, nontrivial transient
regime for supercritical adhesion and nonzero proliferation. The
results of the continuum description in various regimes are
compared with the results of a discrete model \cite{KSS}.

The continuum approach we used is phenomenological. It is
presently unknown how to proceed from the microscopic lattice
model to a macroscopic continuum description even without
proliferation. One way is to use the mean field approximation
\cite{review}, which neglects fluctuations. However, even the
statics, namely the phase diagram, Fig.~\ref{threshold}, in the
mean field approximation is very different from the exact
solution. We have taken another approach, choosing free energy
functional that includes exact statics in it. This allows us to
compare the results of continuum simulations with the results of
discrete model.

The results of continuum theory are in a qualitative agreement
with discrete simulations in a wide region of parameters, in
particular for the velocity of front propagation in subcritical
regime. The continuum approach reproduced a secondary density peak
formation in the transient regime. However, there are important
quantitative differences. Discrete simulations show that the high
density part of the profile diffuses much more slowly than the low
density part. At later times this leads to much smoother fronts
than in the continuum simulations. There is no symmetry between
particles and holes (occupied and empty sites) in the discrete
model \cite{review}. We could introduce a density-dependent
mobility to take this effect into account in the continuum
description. For example, one can consider the following
expression for mobility: $M = 1-q\,u^2$, or similar forms where
$M$ is a decreasing function of density \cite{mobility}.

The modified Cahn-Hilliard equation admits solutions in the form
of propagating fronts. We postulated that the velocity selection
procedure is similar to that of the FK equation and found a
critical velocity, Eq.~(\ref{velocity}). Numerical simulations of
the time-dependent Eq.~(\ref{Cahn}) confirmed this result.
However, the expression for critical velocity becomes invalid in
some region of parameters. Nevertheless, demanding $u \geq 0$ we
can still solve Eq.~(\ref{Cahn}) numerically and observe
propagating fronts with a well-defined velocity, see
Fig.~\ref{comparison}. This problem still needs to be clarified.

An interesting avenue of future work is applying the modified
Cahn-Hilliard equation to the problem of cluster nucleation and
growth in a two-dimensional system to model clustering of
malignant brain tumor cells.

\begin{acknowledgments}
We thank P. Smereka and J. Lowengrub for useful discussions and the
NIH Bioengineering Research
Partnership grant, R01 CA085139-01A2 for support.
\end{acknowledgments}


\begin{thebibliography}{}
\bibitem{fisher2} Note however that the front velocity in the discrete system is smaller than its continuum analog
and approaches it extremely slow \cite{FKPP,FKPP2}.
\end{thebibliography}
\end{document}